\documentclass[aps,prl,showpacs,twocolumn,groupedaddress]{revtex4}
\usepackage{epsfig,amssymb,amsmath}
\newcommand{\zuv}{\hat{\mathbf{z}}} 

\newcommand{\proj}{{\cal P}}
\newcommand{\ket}[1]{\left|#1\right\rangle}  
\newcommand{\br}[1]{\mathbf{r}_{#1}}
\newcommand{\eref}[1]{(\ref{#1})}
\newcommand{\bE}[1]{\mathbf{E}\left(#1\right)}

\newcommand{\bEtot}[1]{\mathbf{E}^{tot}\left(#1\right)}
\newcommand{\bV}[1]{V\left(#1\right)}
\newcommand{\intd}[1]{{\rm d}\mathbf{#1}}
\newcommand{\sumneq}[2]{\substack{#1,#2\\#1\neq#2}}

\begin{document}
\author{J N Kriel and F G Scholtz}
\title{Mapping interacting onto non-interacting quantum Hall systems}
\affiliation{Institute of Theoretical Physics, University of Stellenbosch, South Africa}
\date{\today}
\begin{abstract}
We construct an explicit duality between the interacting quantum Hall system in the lowest Landau level and a non-interacting Landau problem. This is done by absorbing the interaction into the gauge field in the form of an effective magnetic vector potential. The result is analogous to, and illuminates the microscopic origin of, the well-known composite fermion model, but has several advantageous properties. Using this duality we derive, for an arbitrary short-range interaction, analytic expressions for the ground state energy and the excitation gap as functions of the filling fraction. We find good agreement with existing results.
\end{abstract}
\bibliographystyle{prsty}
\maketitle
Our theoretical understanding of the fractional quantum hall effect is based on a number of simple, but very successful qualitative descriptions. The most prominent of these are the Laughlin trial wave-functions \cite{laughlin83} and the composite fermion picture of Jain \cite{jain89}. Despite these successes there is, apart from the work of Murthy and Shankar \cite{murthy03}, no true microscopic derivation of these descriptions. Here we attempt to construct such a dual description explicitly.

We consider a system of $N$ spinless electrons moving in a two dimensional region with surface area $A$. The particles experience an uniform magnetic field perpendicular to the plane and interact via a repulsive two-body potential $V(r)$. The Hamiltonian reads
\begin{equation}
	H=\frac{1}{2m} \sum_i{\left(\mathbf{p}_i-\frac{e}{c}\mathbf{A}_i\right)^2}+V_0\sum_{i<j}\bV{r_{ij}}
	\label{fullh}
\end{equation}
where $e=-\left|e\right|$ is the charge and $m$ the mass of an electron. Interparticle distances are denoted by $r_{ij}=\left|\br{ij}\right|=\left|\br{i}-\br{j}\right|$. In the symmetric gauge the vector potential for an uniform magnetic field $B\zuv$ is $\mathbf{A}_i=(B/2)\zuv\times\mathbf{r}_i$. It is convenient to treat $\bV{r_{ij}}$ as dimensionless and let $V_0$ have the dimension of energy. The kinetic energy term in $H$ is a regular Landau problem which can be solved exactly \cite{laughlin83no2} and  produces the well-known picture of degenerate Landau levels separated by a gap of $\hbar \omega_c$ where $\omega_c=\left(\left|e\right|B\right)/\left(mc\right)$. Let $\phi_0=hc/\left|e\right|$ and $a=\sqrt{\left(\hbar c\right)/\left(|e|B\right)}$  denote the flux quantum and magnetic length respectively. The degeneracy of each Landau level equals $AB/\phi_0$ and the filling fraction $\nu=N\phi_0/AB$ is the, typically fractional, number of levels filled by the electrons. In what follows $\nu$ will always denote the filling fraction of the original Hamiltonian in \eref{fullh}. Although it does not appear explicitly in $H$, we assume that a neutralizing background charge is present in the system.

Following the literature we restrict ourselves to the lowest Landau level (LLL). When $H$ is projected onto this subspace the kinetic energy is a trivial constant, leaving only the projection of the interaction:
\begin{equation}
	\proj\sum_{i<j}\bV{r_{ij}}\proj.
	\label{projh1}
\end{equation}
Later we will encounter Landau problems where singular flux attachments \cite{wilczek82} have been made to the particles. We use $H_0\left(B+\gamma\uparrow\right)$ to denote such a Hamiltonian with an uniform magnetic field $B\zuv$ and flux attachments that carry $\gamma$ flux quanta each.

Now we introduce a new Hamiltonian $\tilde{H}$, which is a modified Landau problem with an additional term added to the vector potential. The form of $\tilde{H}$ is
\begin{equation}
	\tilde{H}=\frac{1}{2m} \sum_i{\left(\mathbf{p}_i-\frac{e}{c}\mathbf{A}_i-\frac{e}{c}\mathbf{\mathcal{A}}_i\right)^2}
	\label{hnop} 
\end{equation}
where $\mathcal{A}_i=\zuv\times\bar{\gamma}\mathbf{E}^{tot}\left(\mathbf{r}_i\right)$ and
\begin{equation}
\mathbf{E}^{tot}\left(\mathbf{r}_i\right)=\sum_{j\neq i} \mathbf{E}\left(\br{ij}\right)=\sum_{j\neq i} E\left(r_{ij}\right)\hat{\mathbf{r}}_{ij}.
\end{equation}
Here $\bar{\gamma}=(\gamma\phi_0)/(2\pi a)$ and $E\left(r\right)$ is a scalar function. Note that both $\mathbf{E}^{tot}\left(\br{}\right)$ and the parameter $\gamma$ are dimensionless. Let us consider the magnetic field corresponding to $\mathcal{A}_i$. The magnetic field at $\br{i}$ due to  $\zuv\times\bar{\gamma}\mathbf{E}\left(\br{ij}\right)$ is
\begin{equation}
B\left(r_{ij}\right)=\zuv\cdot\left[\nabla\times\left(\zuv\times\bar{\gamma}\mathbf{E}\left(\br{ij}\right)\right)\right]=\bar{\gamma}\nabla\cdot\mathbf{E}\left(\br{ij}\right),
\label{bdele}
\end{equation}
while the total field at $\br{i}$ generated by $\mathcal{A}_i$ is
\begin{equation}
B^{tot}\left(\br{i}\right)=\bar{\gamma}\nabla\cdot\mathbf{E}^{tot}\left(\br{i}\right)=\sum_{j\neq i}B\left(r_{ij}\right).
\end{equation}
We interpret this by saying that $\mathcal{A}_i$ attaches a magnetic field to each particle. The radial profile of this field is determined by $E(r)$, e.g. when $E(r)=1/r$ these are the familiar singular flux attachments \cite{wilczek82}.

We will show that $\proj\tilde{H}\proj$ is closely related to the projected interaction in \eref{projh1} when $\mathcal{A}_i$ is defined according to
\begin{equation}
	E\left(r\right)\hat{\mathbf{r}}=-a\nabla V\left(r\right).
	\label{edelv}
\end{equation}
For short-range interactions $E(r)$ will decay quickly at large $r$. This implies that the total flux associated with $\mathcal{A}_i$ is zero, since by Stokes' theorem \mbox{$\int\mathbf{B}\cdot\intd{a}\sim E(r)r$} and $E(r)r\rightarrow0$ as \mbox{$r\rightarrow\infty$}. When the particles are well separated $\mathcal{A}_i$ will therefore have no effect. This reflects the short-range nature of the interactions which $\mathcal{A}_i$ was constructed to mimic.

In order to perform some necessary simplifications we will use a mean-field approximation based on an uniform particle density $\bar{\rho}=N/A=\nu/(2\pi a^2)$. Due to the interactions and fermion statistics the particles are strongly correlated, and we incorporate this in our approximation using the following simple picture. Imagine that around each particle there is a region of radius $r_e$ in which the probability of finding another particle is very small, taken in our case to be zero. The notion that each particle occupies a certain minimum area is supported by the fact that within the LLL the projected spacial coordinates of a particle no longer commute \cite{macris02}. This is implemented in the mean-field approximation by taking the particle density, as seen by a specific particle at $\br{i}$, to be
\begin{equation}
	\rho\left(\mathbf{r}\right)\equiv\left\{\begin{array}{cl}
                0 &\ \ \ \left|\mathbf{r}-\br{i}\right|<r_e\\
                \bar{\rho} &\ \ \ {\rm otherwise}
       \end{array} \right.
   \label{rhodef}
\end{equation}
i.e. a uniform density beginning at a distance $r_e$ from $\br{i}$. The excluded region around $\br{i}$ is  denoted by $A_e^i$. To fix $r_e$ we argue that at low energy the system would tend to maximise the interparticle separation. This effectively assigns an area of $A/N=2\pi a^2/\nu$ to each particle. Taking this region to be a disk of radius $r_e$ we conclude that
\begin{equation}
	\frac{r_e}{a}=\sqrt{\frac{2}{\nu}}.
	\label{redef}
\end{equation}
We remark that the precise value of $r_e$ is not crucial, and does not affect the duality's qualitative results.

One application of this scheme is to approximate the attached fields by a single uniform field. For physical  interactions $V(r)$ should decay smoothly at $r>r_e$. We expect $B(r)$ to share this property, which lends itself to a mean-field approximation:
\begin{eqnarray}
	B^{tot}\left(\br{i}\right)=\sum_{j\neq i}B\left(r_{ij}\right)&\approx&\bar{\rho}\int_{A-A^i_e}{\rm d}\mathbf{r}B\left(\left|\mathbf{r}_i-\mathbf{r}\right|\right) \nonumber \\
&=&-\nu \gamma B\lambda
\label{btotmf2}
\end{eqnarray}
where $\lambda=r_e E(r_e)/a$ is an interaction dependent parameter. The integral over $A$ was dropped since $B(r)$ carries no net flux. The final expression follows from applying Stokes' theorem to the integral over $A_e^i$.

Now we turn to the main task of projecting $\tilde{H}$ onto the LLL. Let $\proj$ denote the projection operator onto the subspace spanned by Slater-determinants formed using the LLL states of $H_0(B)$. Take $\proj\tilde{H}\proj$ and multiply out the $\mathcal{A}_i$ term to obtain
\begin{subequations}
\label{php1}
\begin{eqnarray}
	E_0^{-1}\proj \tilde{H}\proj&=&E_0^{-1}\proj H_0(B)\proj \label{php1a}\\
	&&+\frac{\gamma}{2}\sum_{\sumneq{i}{j}}\proj\frac{a E\left(r_{ij}\right)}{r_{ij}}\left[4 \hat{L}_{ij}+\frac{r_{ij}^2}{a^2}\right]\proj\ \ \ \ \ \ \label{php1b}\\
	&&+{\gamma}^2\sum_i\proj \mathbf{E}^{tot}\left(\mathbf{r}_i\right)\cdot\mathbf{E}^{tot}\left(\mathbf{r}_i\right)\proj \label{php1c}
\end{eqnarray}
\end{subequations}
where $E_0=\hbar\omega_c/2$ and $\hat{L}_{ij}=(2\hbar)^{-1}\zuv\cdot\left[\mathbf{r}_{ij}\times\mathbf{p}_{ij}\right]$ is the $z$ component of the dimensionless relative angular momentum operator. Since $H_0(B)$ assumes a constant value of $N E_0$ within its LLL the first term \eref{php1a} is simply $N$. The remaining two terms are treated separately, starting with \eref{php1b}.
Here we use the fact that any operator containing only relative coordinates is completely characterized in the LLL by its matrix elements between states of the same relative angular momentum. These are the Haldane pseudopotentials \cite{haldane83} and the relevant states are  $\ket{n}=\left(z_i-z_j\right)^n\exp\left[-{r_i}^2/4a^2-{r_j}^2/4a^2\right]$ with $z_k=x_k-iy_k$ and $\hat{L}_{ij}\ket{n}=-n\ket{n}$. Note that fermion statistics require $n$ to be an odd positive integer. It is straightforward to check that \eref{php1b} has the same pseudopotentials as, and is therefore equivalent to, $(2\pi a^2/\phi_0)\sum_{i}\proj B^{tot}\left(\br{i}\right)\proj$. We will approximate $B^{tot}\left(\br{i}\right)$ by its mean-field value derived in \eref{btotmf2}. This reduces \eref{php1b} to a simple scalar: $-\nu\gamma \lambda N$.

Next consider line \eref{php1c} and rewrite the sum as
\begin{equation}
	\sum_i\left[\bEtot{\br{i}}\right]^2=\sum_{j,k}\left[\sum_i\bE{\br{i}-\br{j}}\cdot\bE{\br{i}-\br{k}}\right].\ \ 
\end{equation}
We will use the mean-field approximation to replace the sum over $i$ by an integral. Following \eref{rhodef} the regions around $\br{j}$ and $\br{k}$ are excluded from the domain. The expression above then becomes
\begin{equation}
	\bar{\rho}\sum_{j,k}\int_{A-A^j_e-A_e^k}\intd{r}\,\bE{\br{}-\br{j}}\cdot\bE{\br{}-\br{k}}.
	\label{esqintegral}
\end{equation}
\begin{widetext}
Next we rewrite the dot product using the identity
$\bE{\br{}-\br{j}}\cdot\bE{\br{}-\br{k}}=-\nabla\cdot\left[a\bV{\br{}-\br{j}}\bE{\br{}-\br{k}}\right]\nonumber\\
+a\bV{\br{}-\br{j}}\nabla\cdot\bE{\br{}-\br{k}}$ and then apply Stokes' theorem to the integrals of total divergences. This leads to
\begin{eqnarray}
	\sum_i\left[\bEtot{\br{i}}\right]^2&\approx&-a\bar{\rho}\oint_{\partial A}{\rm d}\mathbf{a}\cdot\left[V^{tot}\left(\mathbf{r}\right)\mathbf{E}^{tot}\left(\mathbf{r}\right)\right]+\frac{\nu}{\gamma\phi_0}\int_{A}\intd{r}\tilde{V}^{tot}\left(\br{}\right)\tilde{B}^{tot}\left(\br{}\right)\nonumber\\
	&+&\sum_{\sumneq{j}{k}}\left[a\bar{\rho}\oint_{\partial A_e^j}{\rm d}\mathbf{a}\cdot\left[V\left(\br{}-\br{j}\right)\mathbf{E}\left(\br{}-\br{k}\right)\right]+a\bar{\rho}\oint_{\partial A_e^k}{\rm d}\mathbf{a}\cdot\left[V\left(\br{}-\br{j}\right)\mathbf{E}\left(\br{}-\br{k}\right)\right]\right]. \label{hoofesq}
\end{eqnarray}
\end{widetext}
Here $\tilde{V}^{tot}\left(\br{}\right)$ and $\tilde{B}^{tot}\left(\br{}\right)$ are the total potential and magnetic field at $\br{}$ due to particles which are at least a distance of $r_e$ from $\br{}$. We will consider each term in \eref{hoofesq} separately. The integral over $\partial A$ may be neglected based on the following electrostatic analogy. Equations \eref{edelv} and \eref{bdele} suggest that we may consider $\mathbf{E}^{tot}\left(\mathbf{r}\right)$ as the electric field due to a charge density distribution proportional to $B^{tot}\left(\mathbf{r}\right)$. Since $B^{tot}\left(\mathbf{r}\right)$ carries no net flux the total charge in this analogy is zero. The integral over $\partial A$ therefore involves the electric field at the edge of a rotationally invariant system with zero net charge. By Gauss' law  $\mathbf{E}^{tot}\left(\mathbf{r}\right)$ can then be taken as zero. Corrections that arise due to boundary effects will not scale extensively and is neglible in the thermodynamic limit. The integral over $\partial A_e^j$ is proportional to $V(r_e)$, which we can choose as zero through a trivial shift in $V(r)$. This is also why no $j=k$ terms appear in the double sum in \eref{hoofesq}. For the integral over $\partial A_e^k$ we approximate $\bV{\br{}-\br{1}}$ by expanding it around $\br{}=\br{k}$ as $\bV{\br{}-\br{j}}=V\left(r_{jk}\right)+\bE{\br{jk}}\cdot\left(\br{}-\br{k}\right)/a$. The second term of the expansion does not contribute to the integral, leaving $\nu\lambda\bV{r_{jk}}$. Finally consider the integral over $A$. Since $\tilde{B}^{tot}\left(\br{}\right)$ only contains contributions from particles lying at least $r_e$ from $\br{}$ it has the same mean-field value as $B^{tot}\left(\br{i}\right)$ in \eref{btotmf2}, even though $\br{}$ is not a particle coordinate. The mean-field value of $\tilde{V}^{tot}(\br{})$ is $V_{bg}\left(\br{}\right)-2V_e$ where \mbox{$V_e=\pi\bar{\rho}\int_0^{r_e}{\rm d}r\,r\,V(r)$} is a constant and $V_{bg}\left(\br{}\right)=\bar{\rho}\int_A\intd{r'}\,V\left(\br{}'-\br{}\right)$ has the form of a potential due to a neutralizing background charge. We now expand the argument of the integral over $A$ to linear order in the fluctuations around these mean-field values. The resulting expression can be treated using approximations very similar to those already encountered. Combining the results of this discussion leads to
\begin{equation}
\sum_i\left[\bEtot{\br{i}}\right]^2=2\nu\lambda\left[H_{int}+NV_e\right],
\end{equation}
where $H_{int}$ contains the particle-particle, particle-background and background-background interactions:
\begin{equation}
	H_{int}=\sum_{i<j}\bV{r_{ij}}-\sum_i V_{bg}\left(\br{i}\right)+\frac{\bar{\rho}}{2}\int\intd{r}\;V_{bg}\left(\br{}\right).
\end{equation}
Returning to \eref{php1} and replacing each line by its simplified form gives
\begin{equation}
	\proj\left[\frac{\tilde{H}}{E_0}-\alpha' N\right]\proj=2\nu\lambda{\gamma}^2 \proj\left[H_{int}+NV_e\right]\proj
	\label{php2}
\end{equation}
where $\alpha'=1-\nu\gamma\lambda$. This is the desired relationship between $\proj\tilde{H}\proj$ and the projected interaction. To exploit this result we need to solve $\proj\tilde{H}\proj$, which we do using a mean-field approach.

First we turn $\tilde{H}$ into a free particle Landau problem by replacing the attached fields by an effective uniform field with strength $-\nu\gamma B\lambda$, as shown in \eref{btotmf2}. This renormalizes the existing field $B\zuv$ by a factor of $\alpha'<1$. The flux in the excluded region $A_e^i$ around $\br{i}$ due to the field attached to particle $i$ cannot be probed directly by other particles. It does however affect the Aharonov-Bohm phases of these particles. It follows that this flux acts as an effective flux attachment carrying $\lambda\gamma$ flux quanta. The mean-field approximation of $\tilde{H}$ now becomes  $H_0\left(\alpha'B+\gamma'\uparrow\right)$ where $\alpha'=1-\nu\gamma\lambda$ and $\gamma'=\gamma\lambda$. The relation $\alpha'=1-\nu\gamma'$ ensures that the total flux present in $\tilde{H}$ and $H_0\left(\alpha'B+\gamma'\uparrow\right)$ are the equal. Upon replacing $\tilde{H}$ by $H_0\left(\alpha' B+\gamma'\uparrow\right)$ equation \eref{php2} becomes strongly reminiscent of the composite fermion picture \cite{jain89,jain92} in that it relates a Landau problem with a weakened magnetic field plus flux attachments to an interacting quantum Hall system. For even integer values of $\gamma'$ the Hamiltonian $H_0\left(\alpha' B+\gamma'\uparrow\right)$ is exactly solvable, although its projection is generally not. However, if the low energy states of  $H_0\left(\alpha' B+\gamma'\uparrow\right)$ lie predominantly within the LLL of $H_0\left(B\right)$, we may drop the projection operators on the left of \eref{php2}. We argue that this is indeed the case. This can be understood intuitively by observing that if the average flux density in $H_0\left(B\right)$ and $H_0\left(\alpha' B+\gamma'\uparrow\right)$ are equal, states which are localised on a radius $r$ will enclose, on average, the same amount of flux and therefore have the same Arahanov-Bohm phase. This will maximise the overlap between the two sets of states. Our construction already satisfies the requirements of this argument. First, it follows from $\alpha'=1-\nu\gamma'$ that the total flux in $H_0\left(B\right)$ and $H_0\left(\alpha' B+\gamma'\uparrow\right)$ are equal. Secondly, the uniform particle density ensures that the average density of singular flux attachments are also uniform, implying that the systems have an equal flux density. It follows that for low energy states with a uniform density, the projection does not dramatically affect the spectrum. The gap in particular should be preserved. We drop the projection operators on the left of \eref{php2} and obtain the final form of the duality:
\begin{equation}
	\frac{H_0\left(\alpha' B+\gamma'\uparrow\right)}{E_0}-\alpha' N=\frac{2\nu{\gamma'}^2}{\lambda}\proj\left[H_{int}+NV_e\right]\proj.
	\label{php3}
\end{equation}
The free parameter $\gamma$ was originally introduced in \eref{hnop}. Its renormalized value $\gamma'=\lambda\gamma$  will be chosen as in the composite fermion literature, where the main consideration is the removal of ground-state degeneracies \cite{jain92}. For the famous series of filling fractions $\nu=(2k+1/p)^{-1}$ we set $\gamma'=2k$, in which case the filling fraction of $H_0\left(\alpha' B+\gamma'\uparrow\right)$ is $\nu_0=1/p$. The flux attachments do not alter the statistics of the particles, and the ground state corresponds to filling the lowest $p$ Landau levels. The finite gap at this filling reveals the incompressibility of the quantum Hall ground state. Expressions for the ground state energy $E_g$ and excitation gap $\Delta$ of $\proj H_{int}\proj$ can easily be read off from \eref{php3} as
\begin{equation}
	\frac{E_g}{N}=-V_e+\frac{(p-1)\lambda}{8pk^2}\label{gseeq}
\end{equation}
and
\begin{equation}
	\Delta=\frac{\lambda}{4k^2p}=\frac{\lambda\nu^2p}{\left(\nu-p\right)^2}. \label{gapeq}
\end{equation}
We now identify the origin of the FQHE's robustness with respect to changes in the form of $V(r)$: By absorbing the interactions into the gauge potential in the form of attached fields a mean-field treatment is possible in which all information about the precise form of $V(r)$ is lost. The interaction dependency only enters through $\lambda$ which rescales the spectrum but cannot affect the existence of the gap. 
\begin{table}[t]
\begin{center}
\begin{tabular}{llllll}
\hline
\hline
$\nu$ & $1/3$ & $2/5$ & $3/7$ & $4/9$ & $5/11$\\
\hline
$\Delta_{1/r}$ \eref{coulombgap} & $0.10206$ & $0.0559$ & $0.0385$ & $0.0295$ & $0.0238$\\
MDD \cite{morf02} & $0.1012$ & $0.05$ & $0.035$ & $0.027$ & $-$\\
CFT \cite{halperin96} & $0.1005$	&	$0.0549$ & $0.0371$ & $0.0276$ & $0.0219$\\
\hline
\hline
\end{tabular}
\\
\vspace{0.2cm}
\begin{tabular}{llll}
\hline
\hline
$\nu$ & $5/11$ & $6/13$ & $7/15$\\
\hline
$\Delta_{1/r}$ \eref{coulombgap}& $0.0238$ & $0.02$ & $0.0173$\\
SLJ \cite{scarola02} & $0.0219(30)$ & $0.0225(41)$ & $0.018(11)$\\
\hline
\hline
\end{tabular}
\end{center}
\caption{The gaps for the Coulomb interaction from \eref{coulombgap} and \cite{morf02,halperin96,scarola02}.
}
\label{coulombtable}
\end{table}
To test the accuracy of \eref{gseeq} and \eref{gapeq}, we consider some specific examples. The ground state energy per particle for the Coulomb interaction ($a/r$) is given by
\begin{equation}
	\frac{E_g}{N}=\sqrt{\frac{\nu}{2}}\left[\frac{p-1}{8pk^2}-1\right],
	\label{coulombgse}
\end{equation}
while the excitation gaps for the $a^n/r^n$ and $a\exp\left[r/(a\kappa)\right]/r$ (Yukawa)  interactions are
\begin{eqnarray}
\Delta_{1/r^n}&=&\frac{n\nu^2p}{\left(\nu-p\right)^2}\left(\frac{\nu}{2}\right)^{n/2}
	\label{rsqgap} \label{coulombgap}\\
\Delta_Y&=&\frac{e^{-\frac{r_e}{a\kappa}}}{4pk^2}\left[\frac{a}{r_e}+\frac{1}{\kappa}\right].
	\label{yukawagap}
\end{eqnarray}
Tables \ref{coulombtable} to \ref{rsqtable} compare our analytic results to those obtained through various other schemes. These include numerical diagonalization as in MDD \cite{morf02}, and the composite fermion theory (CFT) developed in \cite{stern95,halperin96,halperin93}. \mbox{PMJ \cite{park99}}, JK \cite{jain97} and SLJ \cite{scarola02} made use of Monte Carlo methods. The numerical values quoted for PMJ are estimates based on the graphs in \cite{park99}. 

For the Coulomb interaction we found good agreement with the results of CFT, MDD and JK with discrepancies of less then $10\%$ for the gaps and $5\%$ for the ground state energy. Our results for the gap at $5/11$, $6/13$ and $7/15$ agree particularly well with those of SLJ. At these filling fractions up to seven composite fermion Landau levels are filled. At some filling fractions the results of PMJ differ significantly from ours for the $1/r^2$ and Yukawa interactions. This may be due to finite size effects or the extrapolation method used in PMJ; an issue which was raised in MDD. Not shown in the tables are results for the Coulomb interaction at $\nu=1/5$ which appear in \cite{jain97,fano86,girvin85} and range from $0.023$ to $0.025$. Equation \eref{coulombgap} gives $0.0198$; a difference of about $20\%$.
\begin{table}[th]
\begin{center}
\begin{tabular}{llllll}
\hline
\hline
$\nu$ & $1/3$ & $2/5$ & $3/7$ & $1/5$ & $2/9$\\
\hline
$E_g/N$ \eref{coulombgse} & $-0.4082$ & $-0.4193$ & $-0.4243$ & $-0.3162$ & $-0.3281$\\
JK \cite{jain97} & $-0.4098$ & $-0.4328$ & $-0.4422$ & $-0.3275$ & $-0.3427$\\
\hline
\hline
\end{tabular}
\end{center}
\caption{The ground state energy per particle for the Coulomb interaction from \eref{coulombgse} and \cite{jain97}.}
\label{coulombgsetable}
\end{table}
\begin{table}[th]
\begin{center}
\begin{tabular}{lllll}
\hline
\hline
$\nu$ & $1/3$ & $2/5$ & $3/7$ & $4/9$\\
\hline
$\Delta_{1/r^2}$ \eref{rsqgap} & $0.0833$ & $0.05$ & $0.0357$ & $0.0278$\\
PMJ \cite{park99} & $0.0842(11)$ & $0.0609(27)$ & $0.0424(45)$ & $0.0257(77)$\\
\hline
\hline
\end{tabular}\\
\vspace{0.2cm}
\begin{tabular}{lllll}
\hline
\hline
$\nu$ & $1/3$ & $2/5$ & $3/7$ & $4/9$\\
\hline
$\Delta_{Y}$ \eref{yukawagap} & $0.0304$ & $0.01933$ & $0.01406$ & $0.011$\\
PMJ \cite{park99} & $0.0271(4)$ & $0.0215(9)$ & $0.0143(15)$ & $0.0102(24)$\\
\hline
\hline
\end{tabular}
\end{center}
\caption{The gaps for the $1/r^2$ (top) and Yukawa (bottom) interactions from \eref{rsqgap}, \eref{yukawagap} and \cite{park99}. We set $\kappa=1$.}
\label{rsqtable}
\end{table}


\begin{thebibliography}{10}

\bibitem{laughlin83}
R.~B. Laughlin, Phys. Rev. Lett. {\bf 50},  1395  (1983).

\bibitem{jain89}
J.~K. Jain, Phys. Rev. Lett. {\bf 63},  199  (1989).

\bibitem{murthy03}
G. Murthy and R. Shankar, Rev. Mod. Phys. {\bf 75},  1101  (2003).

\bibitem{laughlin83no2}
R.~B. Laughlin, Phys. Rev. B {\bf 27},  3383  (1983).

\bibitem{wilczek82}
F. Wilczek, Phys. Rev. Lett. {\bf 49},  957  (1982).

\bibitem{macris02}
N. Macris and S. Ouvry, J. Phys. A {\bf 35},  4477  (2002).

\bibitem{haldane83}
F.~D.~M. Haldane, Phys. Rev. Lett. {\bf 51},  605  (1983).

\bibitem{jain92}
J. Jain, Adv. Phys. {\bf 41},  105  (1992).

\bibitem{morf02}
R.~H. Morf, N. d\char39{}Ambrumenil, and S. Das~Sarma, Phys. Rev. B {\bf 66},
  75408  (2002).

\bibitem{halperin96}
B. Halperin,  in {\em Perspectives in Quantum Hall Effects}, edited by S. das
  Sarma and A. Pinczuk (Wiley, New York, 1996), p.\ 225.

\bibitem{scarola02}
V.~W. Scarola, S.-Y. Lee, and J.~K. Jain, Phys. Rev. B {\bf 66},  155320
  (2002).

\bibitem{stern95}
A. Stern and B.~I. Halperin, Phys. Rev. B {\bf 52},  5890  (1995).

\bibitem{halperin93}
B.~I. Halperin, P.~A. Lee, and N. Read, Phys. Rev. B {\bf 47},  7312  (1993).

\bibitem{park99}
K. Park, N. Meskini, and J.~K. Jain, J. Phys. C {\bf 11},  7283  (1999).

\bibitem{jain97}
J.~K. Jain and R.~K. Kamilla, Phys. Rev. B {\bf 55},  R4895  (1997).

\bibitem{fano86}
G. Fano, F. Ortolani, and E. Colombo, Phys. Rev. B {\bf 34},  2670  (1986).

\bibitem{girvin85}
S.~M. Girvin, A.~H. MacDonald, and P.~M. Platzman, Phys. Rev. Lett. {\bf 54},
  581  (1985).

\end{thebibliography}
\end{document}